\documentclass{article}

\title{Reduction formalism for Dirac fermions on de Sitter spacetime}
\author{Cosmin Crucean, Radu Racoceanu\\
{\normalsize West University of Timi\c soara, V. Parvan}\\
\normalsize Avenue 4 RO-1900 Timi\c soara, Romania}
\begin{document}
\maketitle

\begin{abstract}
The reduction formulas for Dirac fermions are derived, using the
exact solutions of free Dirac equation on de Sitter spacetime. In
the framework of the perturbation theory one studies the Green
functions and derive the  scatering amplitude in the first orders
of perturbation theory.
\end{abstract}
\section{Introduction}
 The Dirac equation on de Sitter spacetime has been in moving or
 static local charts leading to significant analytical solutions
[2],[9],[10],[11]. The next step is to find solutions for the free
electromagnetic field in moving local charts. The problems that
arises are related to the fact that curved spacetimes have
specific symmetries which , in general, differ from that of
Minkowski spacetime. It is also known that form of the field
equations and their solutions on curved spacetime are strongly
dependent on the tetrad gauge and local chart in which one works
and for that reason we don't expect to find a general solution for
field equations. Also is important to specify that the recent
astrophysical observations shows that the expansion of the
Universe is accelerating and the mathematical model that could
describe the far future limit of our Universe is the de Sitter
model.
\par
 Actually majority of investigations dedicated to Q.E.D on curved spacetimes
don't take into considerations scattering processes. This is
because in the present we don't have one
 scattering theory on curved spacetime. Our aim in this paper is to derive
the reduction formulas for Dirac fermions and to use this result
for developing the scattering theory on de Sitter spacetime. We
will see that the scattering theory on de Sitter spacetime can be
reproduced from that in Minkowski.
\par
We continue in this paper our work related to the developing of
perturbative Q.E.D on de Sitter spacetime. In what follows our
attention will be focused on Dirac field and we shall discuss only
this problem.
 In section 3 we will derive the equations of fields in
interaction and we use this result for constructing the reduction
formalism for Dirac fermions, in section 4 we use this result to
derive the elements of matrix for the amplitude of transition. Our
conclusions is summarized in section 5. The results are presented
in natural units $\hbar =c=1$.
\section{Plane wave}
\par
We start with the exact solutions of the free Dirac equation on de
Sitter spacetime written in [2]. Let us write the de Sitter line
element [1] ,
\begin{equation}
ds^{2}=dt^{2}-e^{2\omega t}d\vec{x}^{2},
\end{equation}
where $\omega$ is the expansion factor. We know that defining a
spinor field on curved spacetime requires one to use the tetrad
fields [1] $e_{\hat{\mu}}(x)$ and $\hat{e}^{\hat{\mu}}(x)$, fixing
the local frames and corresponding coframes which are labelled by
the local indices $\hat{\mu},\hat{\nu},...=0,1,2,3$. The form of
the line element allows one to chose the simple Cartesian gauge
with the non-vanishing tetrad components:
\begin{equation}
e^{0}_{\hat{0}}=e^{-\omega t};
\quad e^{i}_{\hat{j}}=\delta^{i}_{j}e^{-\omega t},
\end{equation}
so that $e_{\hat{\mu}}=e^{\nu}_{\hat{\mu}}e_{\nu}$ and
have the orthonormalization properties
$e_{\hat{\mu}}e_{\hat{\nu}}=\eta_{\hat{\mu}\hat{\nu}},\\
\hat{e}^{\hat{\mu}}e_{\hat{\nu}}=\delta^{\hat{\mu}}_{\hat{\nu}}$
with respect to the Minkowski metric $\eta={\rm
diag}(1,-1,-1,-1)$.
\par
In this gauge the Dirac operator reads [2]:
\begin{equation}
E_{D}=i\gamma^{0}\partial_{t}+ie^{-\omega
t}\gamma^{i}\partial_{i}+\frac{3i \omega}{2}\gamma^{0}.
\end{equation}
\par
Now let us introduce normalized helicity spinors for an arbitrary
vector $\vec{p}$ by notation: $\xi_{\lambda}(\vec{p})$,
\begin{equation}
\vec{\sigma}\vec{p}\xi_{\lambda}(\vec{p})=2p\lambda\xi_{\lambda}(\vec{p}),
\end{equation}
with $\lambda=\pm 1/2$ and where $\vec{\sigma}$ are the Pauli
matrices and $p=\mid\vec{p}\mid$. For witting the solutions of
Dirac equation on de Sitter spacetime we set:
\begin{equation}
k=\frac{m}{\omega},\quad \nu_{\pm}=\frac{1}{2}\pm ik.
\end{equation}
Then the positive frequency modes of momentum $\vec{p}$ and
helicity $\lambda$ that were constructed in [2] as solution of
Dirac equation $E_{D}\psi=m \psi$ using the gamma matrices in
Dirac representation (with diagonal $\gamma^0$) are:
\begin{equation}
U_{\vec{p},\lambda}(t,\vec{x})=\frac{\sqrt{\pi
p/\omega}}{(2\pi)^{3/2}}\left (\begin{array}{c} \frac{1}{2}e^{\pi
k/2}H^{(1)}_{\nu_{-}}(\frac{p}{\omega} e^{-\omega
t})\xi_{\lambda}(\vec{p})\\
\lambda e^{-\pi k/2}H^{(1)}_{\nu_{+}}(\frac{p}{\omega} e^{-\omega
t})\xi_{\lambda}(\vec{p})
\end{array}\right)e^{i\vec{p}\vec{x}-2\omega t},
\end{equation}
where $H^{(1)}_{\nu}(z)$ is the Hankel function of first kind.
\par
Since the charge conjugation  in a curved background is point
independent [8], as in Minkowski case, the negative frequency
modes can be obtained using the charge conjugation,
\begin{equation}
U_{\vec{p},\lambda}(x)\rightarrow
V_{\vec{p},\lambda}(x)=i\gamma^{2}\gamma^{0}(\bar{U}_{\vec{p},\lambda}(x))^{T}\,
\end{equation}
These spinors satisfy the orthonormalization relations [2]:
\begin{eqnarray}
\int d^{3}x
(-g)^{1/2}\bar{U}_{\vec{p},\lambda}(x)\gamma^{0}U_{\vec{p^{\prime}},\lambda^{\prime}}(x)=\\
\nonumber\int d^{3}x
(-g)^{1/2}\bar{V}_{\vec{p},\lambda}(x)\gamma^{0}V_{\vec{p^{\prime}},\lambda^{\prime}}(x)=
\delta_{\lambda\lambda^{\prime}}\delta^{3}(\vec{p}-\vec{p^{\prime}})\\
\nonumber\int d^{3}x
(-g)^{1/2}\bar{U}_{\vec{p},\lambda}(x)\gamma^{0}V_{\vec{p^{\prime}},\lambda^{\prime}}(x)=0,
\end{eqnarray}
where the integration extends on an arbitrary hypersurface
$t=const$ and $(-g)^{1/2}=e^{3\omega t}$. They represent a
complete system of solutions in the sense that [2]:
\begin{equation}
\int d^{3}p
\sum_{\lambda}\left[U_{\vec{p},\lambda}(t,\vec{x})U^{+}_{\vec{p},\lambda}(t,\vec{x^{\prime}})+
V_{\vec{p},\lambda}(t,\vec{x})V^{+}_{\vec{p},\lambda}(t,\vec{x^{\prime}})\right]=e^{-3
\omega t}\delta^{3}(\vec{x}-\vec{x^{\prime}})\,
\end{equation}
\par
The quantization can be done considering the plane wave in
momentum representation [2]:
\begin{eqnarray}
\psi(t,\vec{x})=\psi^{(+)}(t,\vec{x})+\psi^{(-)}(t,\vec{x})\nonumber\\
=\int d^{3}p
\sum_{\lambda}\left[U_{\vec{p},\lambda}(t,\vec{x})a(\vec{p},\lambda)+V_{\vec{p},\lambda}(t,\vec{x})b^{+}(\vec{p},\lambda)\right],
\end{eqnarray}
where the particle $(a,a^{+})$ and antiparticle ($b,b^{+})$
operators satisfy the standard anticommutation relations in
momentum representation:
\begin{equation}
\{a(\vec{p},\lambda),a^{+}(\vec{p^{'}},\lambda^{'})\}=\{b(\vec{p},\lambda),b^{+}(\vec{p^{'}},\lambda^{'})\}=
\delta_{\lambda \lambda^{'}}\delta^{3}(\vec{p}-\vec{p^{'}}).
\end{equation}
From Eq.(9) follows that the equal time anticommutator takes the
canonical form [2]:
\begin{equation}
\{\psi(t,\vec{x}),\bar{\psi}(t,\vec{x^{'}})\}=e^{-3 \omega
t}\gamma^{0}\delta^{3}(\vec{x}-\vec{x^{'}}).
\end{equation}
In any event, these are solutions of the Dirac equation and help
one to write the Green functions in usual manner. Moreover, from
the standard definition of the Feynman propagator:
\begin{eqnarray}
S_{F}(t,t^{'},\vec{x}-\vec{x^{'}})=i\langle0|T[\psi(x)\bar{\psi}(x^{'})]|0\rangle=\theta(t-t^{'})S^{(+)}(t,t^{'},\vec{x}-\vec{x^{'}})\nonumber\\
-\theta(t^{'}-t)S^{(-)}(t,t^{'},\vec{x}-\vec{x^{'}}),
\end{eqnarray}
in [2] was shown that:
\begin{equation}
[E_{D}-m]S_{F}(t,t^{'},\vec{x}-\vec{x^{'}})=-e^{-3 \omega
t}\delta^{4}(x-x^{'}).
\end{equation}
\par Finally we specify that, in general, the partial anticommutator functions
$S^{(+)},S^{(-)}$ are rather complicated since for $t\neq t^{'}$
their time-dependent parts are complicated [2]:
\begin{equation}
S^{(\pm)}(t,t^{'},\vec{x}-\vec{x^{'}})=i\{\psi^{(\pm)}(t,\vec{x}),\bar{\psi}^{(\pm)}(t^{'},\vec{x^{'}})\}.
\end{equation}
With these elements we can try to develop the reduction formalism
and the scattering theory on de Sitter backgrounds.
\section{The reduction formalism}
\par
We start with the interaction between spinors fields and
electromagnetic field on de Sitter spacetime because a theory of
free fields alone has no physical content. The nature of physical
world is revealed to observers only through the interactions
between fields. In this work we use for our calculation the same
formalism as in [3],[5],[6]. We adopt the minimal coupling
corresponding to the classical interaction of a point charge as
the prescription for introducing electrodynamic couplings. Also we
will note the interacting fields with $\psi(x)$ and
$A_{\hat{\mu}}(x)$, where the hated indices indicate label the
components in the local Minkowski frames. This fields will satisfy
one system of equations that can be obtained from an tetrad gauge
invariant action of the free Dirac field, free electromagnetic
field and an interaction term, all minimally coupled with
gravitational field:
\begin{eqnarray}
&&
 \emph{S}[e,\psi,A]=\int d^{4}x
\sqrt{-g}\left\{\frac{i}{2}[\bar{\psi}\gamma^{\hat{\alpha}}D_{\hat{\alpha}}\psi-(\overline{D_{\hat{\alpha}}\psi})\gamma^{\hat{\alpha}}\psi]
\right.\nonumber\\
&&\left.-m\bar{\psi}\psi
-\frac{1}{4}F_{\hat{\alpha}\hat{\beta}}F^{\hat{\alpha}\hat{\beta}}-e
\bar{\psi}\gamma^{\hat{\alpha}}A_{\hat{\alpha}}\psi\right\},
\end{eqnarray}
where $g=|det(g_{\mu\nu})|$, the Dirac matrices
$\gamma^{\hat{\alpha}}$ satisfy
$\{\gamma^{\hat{\alpha}},\gamma^{\hat{\beta}}\}=2
\eta^{\hat{\alpha}\hat{\beta}}$ and covariant derivatives in local
frames,
$D_{\hat{\alpha}}=e^{\mu}_{\hat{\alpha}}D_{\mu}=\hat{\partial}_{\hat{\alpha}}+\hat{\Gamma}_{\hat{\alpha}}$,
are expressed in terms of the spin connections.
\par
The system of equations we obtain from action (16) are:
\begin{eqnarray}
i \gamma^{\hat{\alpha}}D_{\hat{\alpha}}\psi-m\psi=e
\gamma^{\hat{\alpha}}A_{\hat{\alpha}}\psi\nonumber\\
\frac{1}{\sqrt{(-g)}}\partial_{\hat{\alpha}}\left(\sqrt{(-g)}F^{\hat{\alpha}\hat{\beta}}\right)=e\bar{\psi}\gamma^{\hat{\beta}}\psi.
\end{eqnarray}
It is obvious from (17) that in discussing the coupling between
three fields we are up against a nonlinear problem of vast
complexity. This system of equations can be replaced with one
system of integral equations which contain information about
initial conditions. To do this one select two Green functions
$S^{G}(x-y)$ and $D^{G}_{\hat{\alpha}\hat{\beta}}(x-y)$
corresponding to one initial condition, which help us to write the
solutions of system (17) like follows:
\begin{eqnarray}
\psi(x)=\hat{\psi}(x)-e\int
d^{4}y\sqrt{-g}S^{G}(x-y)\gamma^{\hat{\alpha}}A_{\hat{\alpha}}(y)\psi(y)\nonumber\\
A_{\hat{\alpha}}(x)=\hat{A_{\hat{\alpha}}}(x)-e\int
d^{4}y\sqrt{-g}D^{G}_{\hat{\alpha}\hat{\beta}}(x-y)\bar{\psi}(y)\gamma^{\hat{\beta}}\psi(y),
\end{eqnarray}
where $\hat{\psi}(x)$ and $\hat{A_{\hat{\alpha}}}(x)$ are free
fields. One can verify that the first solution from (18) are exact
solution of the first equation from (17), applying $(E_{D}(x)-m)$
to $\psi(x)$, with the observation that
$[E_{D}(x)-m]\hat{\psi}(x)=0$, and using the fact that the Green
function must satisfy:
\begin{equation}
[E_{D}(x)-m]S^{G}(x-y)=-e^{-3 \omega t}\delta^{4}(x-y).
\end{equation}
\par
The Green function for Dirac field will satisfy one equation of
the form (14) and for that reason, $\psi(x)$ is an exact solution
of the system (17).
\par
Equation (18) offers us the possibility of constructing free
fields, which are asymptotic equal (at $t\rightarrow\pm\infty$)
with solutions of system (17). Now we known that the retarded
Green functions $S_{R}(x-y)$ vanishes at $t\rightarrow-\infty$
while the advanced one $S_{A}(x-y)$ vanishes for
$t\rightarrow\infty$. If we write first relation from (18) with
retarded and advanced functions,
\begin{eqnarray}
\psi(x)=\hat{\psi}_{R/A}(x)-e\int
d^{4}y\sqrt{-g}S_{R/A}(x-y)\gamma^{\hat{\alpha}}A_{\hat{\alpha}}(y)\psi(y),
\end{eqnarray}
 then they would define free fields, $\hat{\psi_{R}}(x)$ that satisfy:
\begin{eqnarray}
\lim_{t\rightarrow\mp\infty}(\psi(x)-\hat{\psi}_{R/A}(x))=0.
\end{eqnarray}
The free fields $\hat{\psi_{R}}$ and $\hat{\psi_{A}}$ have mass
$m$ and are equal at $t\pm\infty$ with exact solutions of the
coupled equations and represent the fields before and after the
interaction. Now we known from Minkowski theory that the mass of
the interacting fields may differ from that of free fields because
of the connection between mass and energy, and the energy of
electromagnetic field. In our case we don't know the relation
between mass and energy but it is possible that this difference
between mass of free and interacting fields to be due to the
energy of electromagnetic field and to the coupling with
gravitational field. With this observation the first equation from
(17) can be rewritten:
\begin{equation}
i \gamma^{\hat{\alpha}}D_{\hat{\alpha}}\psi-m\psi=e
\gamma^{\hat{\alpha}}A_{\hat{\alpha}}\psi+\delta m\psi,
\end{equation}
where the difference $\delta m$ will be calculated when one solve
the system of coupled equations.
\par
Like in Minkowski case,the free fields $\hat{\psi_{R}}$ and
$\hat{\psi_{A}}$ are defined up to a normalization constant noted
with $\sqrt{z_{2}}$. Following the same steps like in Minkowski
case one could define $\emph{in}/\emph{out}$ fields:
\begin{eqnarray}
\sqrt{z_{2}}\psi_{in/out}(x)=\psi(x)+e\int
d^{4}y\sqrt{-g}S_{R/A}(x-y)\gamma^{\hat{\alpha}}A_{\hat{\alpha}}(y)\psi(y)\nonumber\\+\delta
m\int d^{4}y\sqrt{-g}S_{R/A}(x-y)\psi(y).
\end{eqnarray}
The $\emph{in}/\emph{out}$ free fields defined above satisfy Dirac
equation, can be written with the help of creation and
annihilation operators and satisfy conditions:
\begin{equation}
\lim_{t\rightarrow\mp\infty}(\psi(x)-\sqrt{z_{2}}\psi_{in/out}(x))=0.
\end{equation}
Using Eq.(22) one could write this fields as follows:
\begin{equation}
\sqrt{z_{2}}\psi_{in/out}(x)=\psi(x)+\int
d^{4}y\sqrt{-g}S_{R/A}(x-y)[E_{D}(y)-m]\psi(y),
\end{equation}
where the Dirac operator reads $E_{D}=i
\gamma^{\hat{\alpha}}D_{\hat{\alpha}}$.
\par
The above normalization rule allows us to write the definition for
the creation and annihilations operators. From (10) if one uses
the orthonormalization relations (8), one obtain:
\begin{eqnarray}
a(\vec{p},\lambda)_{in/out}=\int d^{3}x e^{3\omega
t}\bar{U}_{\vec{p},\lambda}(x)\gamma^{0}\psi_{in/out}(x)\nonumber\\
b^{+}(\vec{p},\lambda)_{in/out}=\int d^{3}x e^{3\omega
t}\bar{V}_{\vec{p},\lambda}(x)\gamma^{0}\psi_{in/out}(x).
\end{eqnarray}
The creation and annihilation operators defined above satisfy the
anticommutation relations (11) and from that it follows that all
the properties of free fields will be preserved.
\par
Before starting our calculations we make a few remarks about the
scattering operator. Denoting the vacuum state $|0\rangle$ , the
one particle states for Dirac fermions can be written:
\begin{eqnarray}
a^{+}(\vec{p},\lambda)_{in/out}|0\rangle=|in/out,1(\vec{p},\lambda)\rangle\nonumber\\
b^{+}(\vec{p},\lambda)_{in/out}|0\rangle=|in/out,\widetilde{1}(\vec{p},\lambda)\rangle.
\end{eqnarray}
If we consider two states $|in,\alpha\rangle$ and
$|out,\beta\rangle$ one define the probability amplitude of
transition from state $\alpha$ to state $\beta$ as the scalar
product of the two states: $\langle out,\beta|in,\alpha\rangle$,
this is just the elements of matrix for scattering operator. This
operator assures the stability of the vacuum state and one
particle state, and in addition transform any $out$ field in the
equivalent $in$ field.
\par
The remaining task is to construct and studies, general matrix
elements which describe the dynamical behavior of interacting
particles. We are interested in the transitions amplitudes for
interacting particles between different initial and final states,
that is the $S$ matrix. As in Minkowski theory we can construct
$n-$particle $in$ and $out$ states as in the free Dirac theory by
repeated application to the vacuum of
$a^{+}(\vec{p},\lambda)_{in/out}$ and
$b^{+}(\vec{p},\lambda)_{in/out}$.
\par
Let us consider the amplitude of one process in which particles
from $in$ states denoted with $\alpha$, together with an electron
$1(\vec{p},\lambda)$ pass in $out$ state, in which we have one
electron $1(\vec{p^{'}},\lambda^{'})$ and particles denoted by
$\beta$. The amplitude of this process can be written as:
\begin{eqnarray}
\langle out \beta,1(\vec{p^{'}},\lambda^{'})|in
\alpha,1(\vec{p},\lambda)\rangle=\langle out
\beta|a_{out}(\vec{p^{'}},\lambda^{'})|in
\alpha,1(\vec{p},\lambda)\rangle\nonumber\\
=\langle out \beta|a_{in}(\vec{p^{'}},\lambda^{'})|in
\alpha,1(\vec{p},\lambda)\rangle+\nonumber\\
\langle out
\beta|(a_{out}(\vec{p^{'}},\lambda^{'})-a_{in}(\vec{p^{'}},\lambda^{'}))|in
\alpha,1(\vec{p},\lambda)\rangle.
\end{eqnarray}
The first term in (28) give if one use the anticommutation
relations (11) the amplitude of one process where the electron
passes from state $in$ in $out$ state without interacting with
particles:
\begin{equation}
\delta_{\lambda\lambda^{'}}\delta(\vec{p}-\vec{p^{'}})\langle out
\beta|in \alpha\rangle.
\end{equation}
It remains to evolve the second terms, and for this we must evolve
the difference:
\begin{equation}
a_{out}(\vec{p^{'}},\lambda^{'})-a_{in}(\vec{p^{'}},\lambda^{'})=\int
d^{3}x e^{3\omega
t}\bar{U}_{\vec{p},\lambda}(x)\gamma^{0}(\psi_{out}(x)-\psi_{in}(x)).
\end{equation}
Using (25) we obtain:
\begin{equation}
\psi_{out}(x)-\psi_{in}(x)=-\frac{1}{\sqrt{z_{2}}}\int
d^{4}y\sqrt{-g} S(x-y)[E_{D}(y)-m]\psi(y),
\end{equation}
where $S(x-y)=S_{R}(x-y)-S_{A}(x-y)$.
 Replacing (30) and (31) in (28) and in addition using:
\begin{equation}
\int d^{3}x e^{3 \omega
t}\bar{U}_{\vec{p},\lambda}(x)\gamma^{0}S(x-y)=i\bar{U}_{\vec{p},\lambda}(y),
\end{equation}
we obtain:
\begin{eqnarray}
\langle out \beta,1(\vec{p^{'}},\lambda^{'})|in
\alpha,1(\vec{p},\lambda)\rangle=\delta_{\lambda\lambda^{'}}\delta^{3}(\vec{p}-\vec{p^{'}})\langle
out \beta|in \alpha\rangle\nonumber\\
-\frac{i}{\sqrt{z_{2}}}\int
\sqrt{-g}\bar{U}_{\vec{p^{'}},\lambda^{'}}(y)[E_{D}(y)-m]\langle
out \beta|\psi(y)|in \alpha,1(\vec{p},\lambda)\rangle d^{4}y.
\end{eqnarray}
\par
The above method can be used to any particle from $in$ or $out$
state. Using the same calculation we find the reduction formula
for one positron from $out$ state:
\begin{eqnarray}
\langle out \beta,\widetilde{1}(\vec{p^{'}},\lambda^{'})|in
\alpha,\widetilde{1}(\vec{p},\lambda)\rangle=\delta_{\lambda\lambda^{'}}\delta^{3}(\vec{p}-\vec{p^{'}})\langle
out \beta|in \alpha\rangle\nonumber\\
+\frac{i}{\sqrt{z_{2}}}\int \sqrt{-g}\langle out
\beta|\bar{\psi}(y)|in
\alpha,\widetilde{1}(\vec{p},\lambda)\rangle
[\bar{E}_{D}(y)-m]
V_{\vec{p^{'}},\lambda^{'}}(y)d^{4}y,
\end{eqnarray}
where we note
$\bar{E}_{D}=-i\gamma^{\hat{\alpha}}\bar{D}_{\hat{\alpha}}$ . The
calculations of reduction formulas for electron and positron from
$in$ state can be done using the above method obtaining for the
transition amplitude:
\begin{eqnarray}
\langle out \beta,1(\vec{p^{'}},\lambda^{'})|in
\alpha,1(\vec{p},\lambda)\rangle=\delta_{\lambda\lambda^{'}}\delta^{3}(\vec{p}-\vec{p^{'}})\langle
out \beta|in \alpha\rangle\nonumber\\
-\frac{i}{\sqrt{z_{2}}}\int \sqrt{-g}\langle out
\beta,1(\vec{p^{'}},\lambda^{'})|\bar{\psi}(y)|in \alpha\rangle
[\bar{E}_{D}(y)-m]U_{\vec{p},\lambda}(y)d^{4}y,\nonumber\\
\langle out \beta,\widetilde{1}(\vec{p^{'}},\lambda^{'})|in
\alpha,\widetilde{1}(\vec{p},\lambda)\rangle=\delta_{\lambda\lambda^{'}}\delta^{3}(\vec{p}-\vec{p^{'}})\langle
out \beta|in \alpha\rangle\nonumber\\
+\frac{i}{\sqrt{z_{2}}}\int
\sqrt{-g}\bar{V}_{\vec{p},\lambda}(y)[E_{D}(y)-m]\langle out
\beta,\widetilde{1}(\vec{p^{'}},\lambda^{'})|\psi(y)|in
\alpha\rangle d^{4}y.
\end{eqnarray}
Now we can proceed with the reduction of the second particle. For
this we suppose that we already done the reduction of the first
electron from $out$ state and we obtain (33). Now if one suppose
that in the particles denoted by $\beta$ exist another electron
then $\beta=\beta^{'}+1(\vec{p^{''}},\lambda^{''})$, it follows to
reduce this electron. The matrix element in which we are
interested appears in (33):
\begin{eqnarray}
\langle out \beta|\psi(y)|in
\alpha,1(\vec{p},\lambda)\rangle=\langle out
\beta^{'},1(\vec{p^{''}},\lambda^{''})|\psi(y)|in
\alpha,1(\vec{p},\lambda)\rangle\nonumber\\
=\langle out
\beta^{'}|[a_{out}(\vec{p^{''}},\lambda^{''})\psi(y)+\psi(y)a_{in}(\vec{p^{''}},\lambda^{''})]|in
\alpha,1(\vec{p},\lambda)\rangle\nonumber\\-\langle out
\beta^{'}|\psi(y)a_{in}(\vec{p^{''}},\lambda^{''})|in\alpha,1(\vec{p},\lambda)\rangle.
\end{eqnarray}
The last terms is the form $\langle out
\beta^{'}|\psi(y)|in\alpha\rangle
\delta_{\lambda\lambda^{''}}\delta^{3}(\vec{p^{''}}-\vec{p})$ ,
and corresponds to a process where electron don't interact with
other particles and for that reason this term is not interesting
for us. The first term in (36) give the amplitude that interest
us. We will start with the evaluation of the sum using (26):
\begin{eqnarray}
a_{out}(\vec{p^{''}},\lambda^{''})\psi(y)+\psi(y)a_{in}(\vec{p^{''}},\lambda^{''})&=&\int
e^{3 \omega
t}\bar{U}_{\vec{p^{''}},\lambda^{''}}(x)\gamma^{0}\left[\psi_{out}(x)\psi(y)\right.\nonumber\\
&&\left.+\psi(y)\psi_{in}(x)\right]d^{3}x,
\end{eqnarray}
then using (25) one obtain:
\begin{eqnarray}
&&
a_{out}(\vec{p^{''}},\lambda^{''})\psi(y)+\psi(y)a_{in}(\vec{p^{''}},\lambda^{''})=
\frac{1}{\sqrt{z_{2}}}\int e^{3 \omega
t}\bar{U}_{\vec{p^{''}},\lambda^{''}}(x)\gamma^{0}\left[\psi(x)\psi(y)+\right.\nonumber\\
&&\left.\psi(y)\psi(x)\right]d^{3}x +\frac{1}{\sqrt{z_{2}}}\int
\int e^{6 \omega
t}\bar{U}_{\vec{p^{''}},\lambda^{''}}(x)\gamma^{0}
\left[-\theta(z^{0}-x^{0})S(x-z)[E_{D}(z)-m]\right.\nonumber\\
&&\left.\times\psi(z)\psi(y)+\theta(x^{0}-z^{0})S(x-z)[E_{D}(z)-m]\psi(y)\psi(z)\right]d^{3}x
d^{4}z.
\end{eqnarray}
Using (32) and the explicit form of Dirac operator
$E_{D}=i\gamma^{0}\partial_{t}+ie^{-\omega
t}\gamma^{i}\partial_{i}+\frac{3i \omega}{2}\gamma^{0}$ one
obtain:
\begin{eqnarray}
a_{out}(\vec{p^{''}},\lambda^{''})\psi(y)+\psi(y)a_{in}(\vec{p^{''}},\lambda^{''})=
\frac{1}{\sqrt{z_{2}}}\int e^{3 \omega
t}\bar{U}_{\vec{p^{''}},\lambda^{''}}(x)\gamma^{0}\{\psi(x),\psi(y)\}d^{3}x\nonumber\\
+\frac{i}{\sqrt{z_{2}}}\int e^{3 \omega
t}\bar{U}_{\vec{p^{''}},\lambda^{''}}(z)[E_{D}(z)-m]
\left[-\theta(z^{0}-x^{0})\psi(z)\psi(y)
+\theta(x^{0}-z^{0})\psi(y)\psi(z)\right]d^{4}z\nonumber\\
+\frac{i}{\sqrt{z_{2}}}\int e^{3 \omega
t}\bar{U}_{\vec{p^{''}},\lambda^{''}}(z)\left[i\gamma^{0}\delta(z^{0}-x^{0})\psi(z)\psi(y)
+i\gamma^{0}\delta(z^{0}-x^{0})\psi(y)\psi(z)\right]d^{4}z.
\end{eqnarray}
The last term appears because we place Dirac equation in front of
parenthesis, thus acting on distributions $\theta$ and generating
the last term with changed sign. Integrating after $dz^{0}$ in the
last term from (39) and changing the spatial variables
$\vec{z}\rightarrow\vec{x}$, one obtain just the first term from
(39) with opposite sign. Now observing that in (37) the spatial
integral is done at the time $x^{0}$ which is arbitrary,
 one could choice $x^{0}=y^{0}$ and obtain the
chronological product $T[\psi(z)\psi(y)]$. With this observation
we finally obtain:
\begin{eqnarray}
a_{out}(\vec{p^{''}},\lambda^{''})\psi(y)+\psi(y)a_{in}(\vec{p^{''}},\lambda^{''})=\nonumber\\
-\frac{i}{\sqrt{z_{2}}}\int\sqrt{-g}\bar{U}_{\vec{p^{''}},\lambda^{''}}(z)[E_{D}(z)-m]T[\psi(z)\psi(y)]d^{4}z.
\end{eqnarray}
The matrix element will be:
\begin{eqnarray}
\langle out
\beta^{'}|a_{out}(\vec{p^{''}},\lambda^{''})\psi(y)+\psi(y)a_{in}(\vec{p^{''}},\lambda^{''})|in
\alpha,1(\vec{p},\lambda)\rangle =\nonumber\\
-\frac{i}{\sqrt{z_{2}}}\int\sqrt{-g}\bar{U}_{\vec{p^{''}},\lambda^{''}}(z)[E_{D}(z)-m]\langle
out \beta^{'}|T[\psi(z)\psi(y)]|in
\alpha,1(\vec{p},\lambda)\rangle d^{4}z.
\end{eqnarray}
\par
From the above calculations one sees that the reduction of the
second particle is done using the same method as for the first
particle. In the matrix element that interest us (41), appears two
field operators that are multiplied in chronological order.
Repeating the calculation for other particles from $in$ and $out$
states one observe that the reduction calculus is the same,
indifferent what type of particle is reduced.
\par
Also we can obtain a generalization of the above formulas,
supposing that we have $n-out$ and $m-in$ , Dirac particles, after
we apply the reduction formalism finally obtain:
\begin{eqnarray}
\langle
out,(\vec{p^{'}_{1}},\lambda^{'}_{1}),...,(\vec{p^{'}_{n}},\lambda^{'}_{n})|in,(\vec{p_{1}},\lambda_{1}),...,(\vec{p_{m}},\lambda_{m})\rangle=(-)^{m+n}\nonumber\\
\left(-\frac{i}{\sqrt{z_{2}}}\right)^{m+n}\prod^{n}_{i=1}\int
d^{4}x_{i}\sqrt{-g(x_{i})}\prod^{m}_{j=1}\int
d^{4}y_{j}\sqrt{-g(y_{j})}\bar{U}_{\vec{p^{'}_{i}},\lambda^{'}_{i}}(x_{i})[E_{D}(x_{i})-m]\nonumber\\
\times\langle
0|T[\psi(x_{i})...\psi(x_{n})\bar{\psi}(y_{j})...\bar{\psi}(y_{m})]|0
\rangle[\bar{E}_{D}(y_{j})-m]U_{\vec{p_{j}},\lambda_{j}}(y_{j}).
\end{eqnarray}
The sign $(-)^{m+n}$ is governed by the number of sign changes
dictated by the definition of time ordering for fermion fields.
\par
 Now we are in the position of making important
observations about the above reduction method. One can show that
all particles, will be replaced by formulas that was obtain in the
reduction of one particle. Also when more particles are reduced,
in matrix element appears the time ordered products of
corresponding field operators. Every particle will be replaced
after reduction with expressions which depend on field operator
$\psi(x)$ ((33),(34),(35)). After we reduce all particles from
$in$ and $out$ states we arrive at a vacuum expectation value of
time ordered product of fields. We don't write explicitly the
spinorial indices because is obviously that $E_{D}(y)$ will act as
$(4 \times 4)$ matrix and as differential operator just on spinor
$\psi(y)$. One observe that thought reduction method the
amplitudes was written as function of fundamental solutions of the
free Dirac equation on de Sitter spacetime, and as function of
vacuum expectation value of time ordered product of fields. The
vacuum expectation value of time ordered product of fields,
together with normalization constants ($\sqrt{z_{2}}$) define the
Green functions of the interacting fields. One can associate one
Green function to any process of interaction. After we apply the
reduction formalism, the Green functions of interacting fields
must be calculated.
\par
At the end of this section we write the reduction rules for
particles and antiparticles, denoting one electron by
$1(\vec{p},\lambda)$ and one positron by
$\widetilde{1}(\vec{p},\lambda)$, after reduction of particles
from $in$ and $out$ states one obtain:
\begin{eqnarray}
 1(\vec{p^{'}},\lambda^{'})\quad out \quad\rightarrow
-\frac{i}{\sqrt{z_{2}}}\int(-g)^{1/2}\bar{U}_{\vec{p^{'}},\lambda^{'}}(x)[E_{D}(x)-m]\psi(x)d^{4}x,\nonumber\\
\widetilde{1}(\vec{p^{'}},\lambda^{'})\quad out \quad\rightarrow
\frac{i}{\sqrt{z_{2}}}\int(-g)^{1/2}\bar{\psi}(x)[\bar{E}_{D}(x)-m]V_{\vec{p^{'}},\lambda^{'}}(x)d^{4}x,\nonumber\\
1(\vec{p},\lambda)\quad in \quad\rightarrow
-\frac{i}{\sqrt{z_{2}}}\int(-g)^{1/2}\bar{\psi}(x)[\bar{E}_{D}(x)-m]U_{\vec{p},\lambda}(x)d^{4}x,\nonumber\\
\widetilde{1}(\vec{p},\lambda)\quad in \quad\rightarrow
\frac{i}{\sqrt{z_{2}}}\int(-g)^{1/2}\bar{V}_{\vec{p},\lambda}(x)[E_{D}(x)-m]\psi(x)d^{4}x.
\end{eqnarray}
\section{The perturbation theory}
\par
The Green functions of the interacting fields can't be calculated
exact and for that reason we will use perturbation methods. The
form of the amplitudes obtained from reduction formalism allows
one to use perturbation calculus.
\par
It is clear now that the entirely perturbation theory on de Sitter
spacetime can be reproduced from Minkowski theory [3],[4],[5],[6].
For calculating the Green functions we must write then as
functions of free fields, because we know their form and
properties. We write the Green function in generally as follows:
\begin{equation}
G(y_{1},y_{2},...,y_{n})=
\frac{1}{\langle0|\widetilde{S}|0\rangle}\langle0|T[\hat{\psi}(y_{1})\hat{\psi}(y_{2})...\hat{\psi}(y_{n}),\widetilde{S}]|0\rangle,
\end{equation}
where $\widetilde{S}$ is a unitary operator and have a closer form
with the same operator from Minkowski theory. Like in Minkowski
case the operator $\widetilde{S}$ must be correlated with
scattering operator $S$. One can show that this two operators are
equal up to a phase factor.
\par
Then entire perturbation calculus is based on development of
operator $\widetilde{S}$:
\begin{eqnarray}
\widetilde{S}=T e^{-i\int
\sqrt{-g}\mathcal{L}_{I}(x)d^{4}x}=1+\nonumber\\
\sum\limits_{n=1}^\infty
\frac{(-i)^{n}}{n!}\int(-g)^{n/2}T[\mathcal{L}_{I}(x_{1})...\mathcal{L}_{I}(x_{n})]d^{4}x_{1}...d^{4}x_{n},
\end{eqnarray}
where the density lagrangian of interaction reads:
$\mathcal{L}_{I}(x)=-e
:\bar{\psi}(x)\gamma^{\hat{\alpha}}A_{\hat{\alpha}}(x)\psi(x):-\delta
m :\bar{\psi}(x)\psi(x):$. Each term from (45) corresponds to a
rang from perturbation theory. Replacing (45) in the expression of
Green function (44) one obtain perturbation series which allows
one to calculate the amplitude in any order. The term of rang $n$
of this development is:
\begin{equation}
\frac{(-i)^{n}}{n!\langle0|\widetilde{S}|0\rangle}\int(-g)^{n/2}\langle0|T[\hat{\psi}(y_{1})\hat{\psi}(y_{2})...\hat{\psi}(y_{n}),
{\mathcal{L}}_{I}(x_{1})...{\mathcal{L}}_{I}(x_{n})]|0\rangle
d^{4}x_{1}...d^{4}x_{n}.
\end{equation}
 The evaluation of the integrant from (46) is the same as in
Minkowski case, we have a cinematic part which is obtained from
reduction formalism and a dynamic one represented by operator
$\widetilde{S}$. Following the same steps as in Minkowski theory
one will make the $T$ contractions between cinematic and dynamic
part, with the observation that the $T$ contractions between
fields from cinematic part will not give contributions to the
scattering amplitude. Also in the case of $T$ contractions between
fields from dynamic part we have two possibilities. One is that
all fields from dynamic part coupled fields from cinematic one and
the second is that one part of the fields from dynamic part
contract between them. The second possibility will give one term
of the form $\langle0|\widetilde{S}|0\rangle$, which will simplify
the nominator.
\par
As an application to our formalism we can obtain the scattering
amplitudes in first orders in perturbation theory, thus completing
the framework that one needs for calculating scattering processes
in the first order of perturbation theory, on de Sitter expanding
universe. Now using the reduction formalism developed in section
$3$ for one amplitude of the form $\langle
out,1(\vec{p^{'}},\lambda^{'})|in,1(\vec{p},\lambda)\rangle$ , one
obtain the development:
\begin{eqnarray}
\langle
out,1(\vec{p^{'}},\lambda^{'})|in,1(\vec{p},\lambda)\rangle=\delta_{\lambda\lambda^{'}}\delta^{3}(\vec{p}-\vec{p^{'}})
-\nonumber\\
\frac{1}{z_{2}}\int\int(-g)\bar{U}_{\vec{p^{'}},\lambda^{'}}(y)[E_{D}(y)-m]
\langle0|T[\psi(y)\bar{\psi}(z)]|0\rangle
[\bar{E}_{D}(z)-m]U_{\vec{p},\lambda}(z)d^{4}y d^{4}z.
\end{eqnarray}
 Using (46) and neglecting the first term in (47), one obtain the for
scattering amplitude in first order of perturbation theory:
\begin{eqnarray}
A_{i\rightarrow
f}=\frac{-ie}{\langle0|\widetilde{S}|0\rangle}\int\int\int(-g)^{3/2}\bar{U}_{\vec{p^{'}},\lambda^{'}}(y)[E_{D}(y)-m]
\langle0|T[:\psi(y)\bar{\psi}(z):\nonumber\\:\bar{\psi}(x)\gamma_{\hat{\mu}}A^{\hat{\mu}}(x)\psi(x):]|0\rangle
[\bar{E}_{D}(z)-m] U_{\vec{p},\lambda}(z)d^{4}y d^{4}z d^{4}x.
\end{eqnarray}
After we make the $T$ contractions between cinematic and dynamic
part using the method from Minkowski theory,
 and use (13) and (14), finally obtain:
\begin{equation}
A_{i\rightarrow f}=-ie
\int\sqrt{-g}\bar{U}_{\vec{p^{'}},\lambda^{'}}(x)\gamma_{\hat{\mu}}A^{\hat{\mu}}(x)U_{\vec{p},\lambda}(x)d^{4}x.
\end{equation}
The above expression is just the scattering amplitude that we use
in our previous work [7], for calculate the Coulomb scattering on
de Sitter expanding universe. Following the same steps as above
one can obtain the scattering amplitude in superior orders in
perturbation theory. With the above calculations we establish the
general rules of calculation which can be used in the language of
Feynman graphs.
\section{Conclusion}
\par
In this paper we investigate the reduction formalism for solution
of the free Dirac equation on de Sitter background. We obtain that
the reduction formalism can be calculated using the same method as
in Minkowski case. Also we show that after reduction of particles
from $in$ and $out$ states one obtain the vacuum average of fields
written in chronological order. As in the Minkowski case to any
interaction one can associate one Green function, that can be
evolved using perturbation theory. From our formalism of reduction
and using perturbation theory, we deduce the correct definition
for the scattering amplitudes.
\par
From our point of view is important to do the same calculations
for the electromagnetic field, thus completing this theory and the
framework that one needs for developing perturbative Q.E.D on de
Sitter spacetime.
\par
\textbf{Acknowledgements}
  \par
We would like to thank Professor Ion
I.Cot\u aescu  for encouraging us to do this work and for reading
the manuscript.

\end{document}